\documentclass[12pt]{iopart}

\usepackage{iopams}  

\usepackage[english]{babel}
\usepackage{todonotes}
\usepackage{graphicx}

\usepackage[utf8]{inputenc}
\usepackage{float}
\usepackage{hyperref}

\newcommand{\TT}{\mathcal{T}}
\newcommand{\UU}{\mathcal{U}}
\newcommand{\FF}{\mathcal{F}}
\newcommand{\KK}{\mathcal{K}}
\newcommand{\eqref}[1]{(\ref{#1})}

\begin{document}

\title{Formation of trade networks by economies of scale and product differentiation}

\author{    Chengyuan Han$^{1,2}$\href{https://orcid.org/0000-0001-5220-402X}{\includegraphics[width=3.2mm]{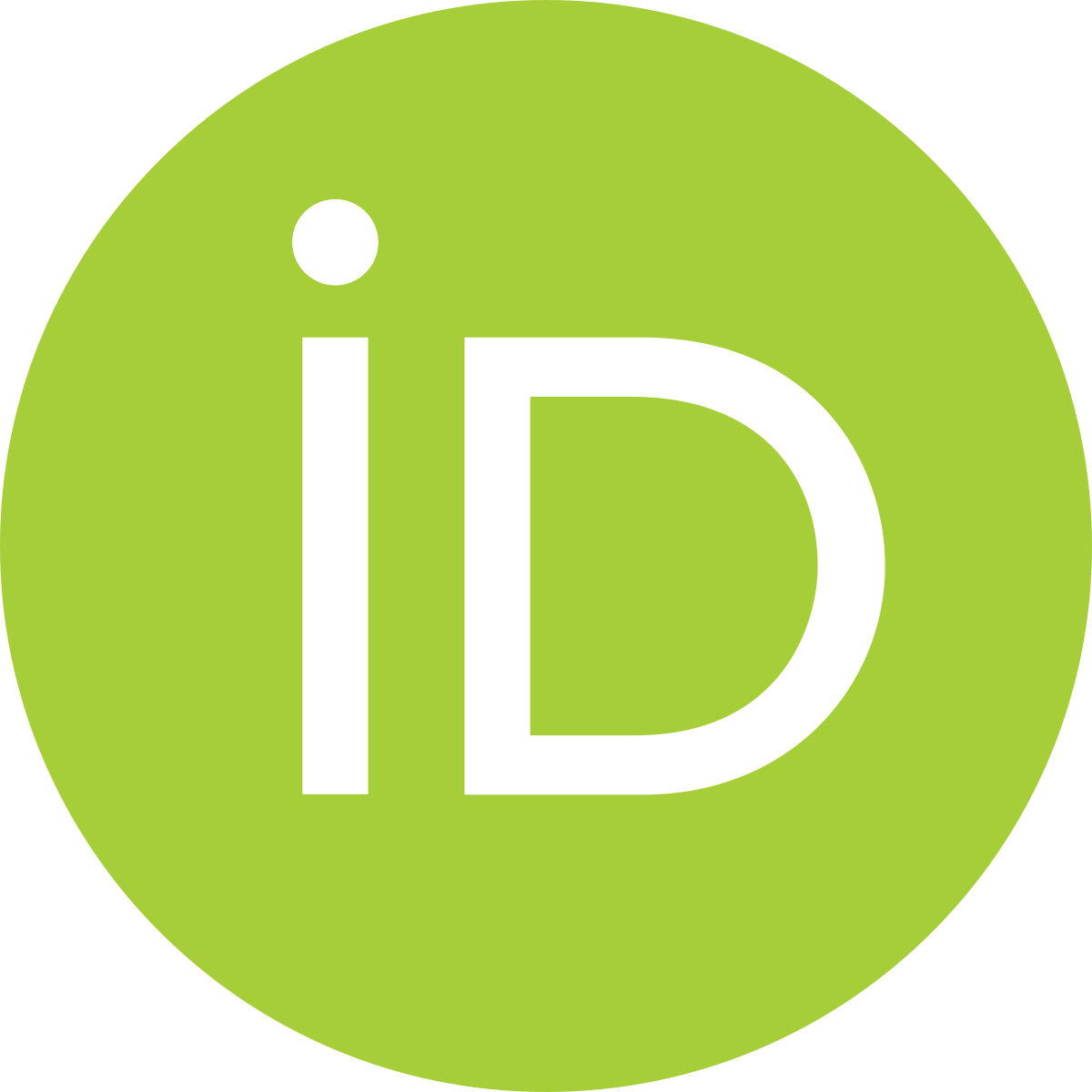}}, 
            Malte Schröder$^{3}$\href{https://orcid.org/0000-0001-8756-9918}{\includegraphics[width=3.2mm]{orcid.png}}, 
            Dirk Witthaut${}^{1,2,4}$\href{https://orcid.org/0000-0002-3623-5341}{\includegraphics[width=3.2mm]{orcid.png}}, and
            Philipp C. Böttcher$^{1,4}$\href{https://orcid.org/0000-0002-3240-0442}{\includegraphics[width=3.2mm]{orcid.png}}
        }

\address{${}^1$Forschungszentrum J\"ulich, 
               Institute of Energy and Climate Research -- Systems Analysis and Technology Evaluation (IEK-STE), 52428 J\"ulich, Germany\\
	       ${}^2$Institute for Theoretical Physics, University of Cologne, K\"oln, 50937, Germany\\ 
	       ${}^3$Center for Advancing Electronics Dresden (cfaed) and Institute for Theoretical Physics, Technische Universität Dresden, Dresden, 01062, Germany \\ 
	       ${}^4$Forschungszentrum J\"ulich, 
               Institute of Energy and Climate Research -- Energy Systems Engineering (IEK-10), 52428 J\"ulich, Germany\\
		
}

\begin{abstract}
Understanding the structure and formation of networks is a central topic in complexity science. 
Economic networks are formed by decisions of individual agents and thus not properly described by established random graph models. 
In this article, we establish a model for the emergence of trade networks that is based on rational decisions of individual agents. 
The model incorporates key drivers for the emergence of trade, comparative advantage and economic scale effects, but also the heterogeneity of agents and the transportation or transaction costs. 
Numerical simulations show three macroscopically different regimes of the emerging trade networks. 
Depending on the specific transportation costs and the heterogeneity of individual preferences, we find centralized production with a star-like trade network, distributed production with all-to-all trading or local production and no trade. 
Using methods from statistical mechanics, we provide an analytic theory of the transitions between these regimes and estimates for critical parameters values.    
\end{abstract}

\section{Introduction}

Trade networks are essential to today's international economy \cite{krugman1992geography,schweitzer2009economic}. 
From supply chains to financial transactions, goods and values are exchanged among almost every region on Earth. 
The importance of economic connectivity and complexity becomes most obvious in case of a disturbance: The world financial crisis of 2007/2008 emerged after inter-bank claims and liabilities had been growing for decades, enabling a rapid spread of financial risks \cite{gai2011complexity, helbing2013globally}. 
Similarly, disruptions of global transportation networks due to the Covid-19 pandemic have led to a substantial loss of production in various regions \cite{guan2020global}. 

The emergence of connectivity and the structure of networks are essential topics in complexity science~\cite{newman2003structure, pastor2003statistical, havlin2012challenges}. 
Traditionally, ensembles of random networks have been used to describe essential features of real-world networks such as the small-world effect or the emergence of hubs~\cite{watts1998collective,barabasi1999emergence}. 
On the one hand, percolation theory enables far-reaching insights into the formation and robustness of ensembles of networks \cite{newman2010networks}. 
On the other hand, optimization models have been used to describe the structure of biological networks \cite{katifori2010damage,kaiser2020discontinuous}, assuming that evolution created structures that provide a certain function in an optimal way. 
Both approaches are of limited use in the modeling of economic networks, where links are neither established at random, nor following a single, global objective. Instead, links are established deliberately on the basis of individual decisions.

The emergence of trade is a central subject of economics. 
In the celebrated Ricardian model, trade patterns are determined by the regional differences in productivity providing a comparative advantage~\cite{krugman2009international}. Initially formulated in terms of productivity of labor, the Ricardian model was later generalized and formulated in terms of opportunity costs~\cite{bernhofen2005gottfried}. Predictions of these trade theories have been tested in various empiric studies, see~\cite{bernhofen2005empirical} for a recent example. A comparative advantage typically arises when the trading countries have strongly different characteristics. Explaining the emergence of trade between similar countries required a substantial extension of the theory. In a landmark study from 1979~\cite{krugman1979increasing}, Krugman pointed out the importance of economies of scale and transportation costs. Economic scale effects foster the centralization of production, as they lead to lower production costs for large producers and thus to advantages in competition. Transportation costs may lead to a home market effect, where regions or countries with a higher demand have an advantage and tend to export a good. Mathematical models in this field often feature the Dixit-Stiglitz model of consumer preferences in terms of a utility function with a constant elasticity of substitution~\cite{dixit1977monopolistic}. Krugman later extended these ideas towards a comprehensive theory of trade and economic geography~\cite{krugman1991increasing}. Notably, economies of scale can also lead to lock-in effects~\cite{krugman1981trade}. A well readable introduction to this topic is provided in the text book~\cite{krugman1992geography}.

However, transportation and production costs are not the only attributes that determine economic interactions and trade. 
Discrete choice theory investigates how economic agents reach a decision on the basis of both observed and unobserved attributes \cite{anderson1992discrete,lancaster1990economics}. 
The agents' preferences vary, in particular with respect to the unobserved attributes, and so do the agents' choices. 
Thus, discrete choice models are intrinsically stochastic, describing the probability of choice on an individual level or the demand for certain goods or brands on an aggregate level. 
As a consequence, populations of heterogeneous consumers demand differentiated goods \cite{anderson1992logit,matvejka2015rational}.

Recent research in statistical mechanics and network science has provided a variety of empirical insights into the structure of international trade networks. Early studies of the structure of trade networks have shown a scaling behaviour \cite{serrano2003topology} and the correlations between different commodities \cite{barigozzi2010multinetwork}. An explanation for the observed scaling has been suggested in \cite{garlaschelli2004fitness} in terms of a fitness model. Changes in the world trade network in the last decades were studied in \cite{karpiarz2014international,he2010structure}, pointing out the importance of (fractal) geography to understand trade patterns and quantifying the robustness to economic shocks.

In this article we establish a model for the formation of trade networks combining essential concepts of economics and statistical physics. 
Trade links are established by the decisions of individual agents taking into account differentiated preferences in the spirit of discrete choice theory.
The cost functions incorporate economic scale effects, making the decision problems nonlinear and interdependent.  
Statistical physics guides the numerical solution of the problem as well as the analysis of the results. 
We compute the `phase diagram' of the emergent trade network using a self-consistent method and provide an approximate analytic theory for the transitions between different regimes of the emerging trade network.
The current article generalizes previous models~\cite{schroder2018hysteretic,han2019winner} which neglected product differentiation and focused solely on percolation aspects.

\section{Models and Methods}
\label{sec:trade-model}

\subsection{Discrete choice Theory}

Discrete choice theory considers individual agents or consumers which may choose from different discrete options. The preferences of the individuals vary leading to product differentiation \cite{lancaster1990economics,anderson1992discrete,anderson1992logit,manski1990structural}

To formalize this model, we consider a set of nodes or vertices $i \in \{1,\ldots,N\}$ representing well-defined spatial units, each inhabited by a large number of agents $D_i$. A single agent $a$ at node $i$ chooses to purchase a good from different nodes $j \in \{1,\ldots,N\}$ at different effective prices $\tilde{p}_{ji}$. However, the price is not the only factor that determines consumer behavior and preferences generally differ. 
Hence, the utility of an individual agent $a$ for a good from node $j$ is 
\begin{equation}
    \UU_a(j) = \UU_0  - \tilde{p}_{ji} + \TT \epsilon_a(j).
    \label{eq:utility-agent}
\end{equation}
with a constant term $\UU_0>0$. It is assume that the utility of a good from node $j$ decreases as the effective price $\tilde{p}_{ji}$ increases which holds for all agents alike. The difference of the individual preferences of the agents $a$ are summarized in the  $\epsilon_a(j)$, which is typically unknown a priori and thus modeled as random variables. The parameter $\TT>0$ measures how strongly preferences vary between individual agents. A common assumption in the economic literature is that the $\epsilon_a(j)$ are independent and identically Gumbel distributed, which leads to the classic multinomial logit model \cite{anderson1992logit,manski1990structural}.
The probability of an agent $a$ at node $i$ choosing alternative $j$ is then given by 
\begin{equation}
    \mathcal{P}_a(j) = \frac{\exp(- \tilde{p}_{ji}/\TT)} 
    {\sum_{\ell=1}^N\exp(- \tilde{p}_{\ell i}/\TT)} .
\end{equation}

Now consider the cumulative purchases $S_{ji}$ made by all agents at node $i$ from all nodes $j \in \{1,\ldots,N\}$. Assuming that the number of agents $D_i$ at node $i$ is sufficiently large, we can replace the amount by its expected value and obtain
\begin{equation}
    S_{ji} = D_i \, \frac{\exp(- \tilde{p}_{ji}/\TT)} 
    {\sum_{\ell=1}^N\exp(- \tilde{p}_{\ell i}/\TT)} .
    \label{eq:skifull}
\end{equation}
This expression is the starting point of our analysis. In the limiting case $\TT \rightarrow 0$, differences between the agents at a node $i$ vanish and all agents purchase from the same node $j = i^*$ for which the prices are smallest,
\begin{equation}
    S_{ji} = \left\{ \begin{array}{l l l}
        D_i & \, \mbox{for} \, & j = i^* \\
        0   & & j \neq i^* ,
    \end{array}
    \right.
    \label{eq:Sji-Tzero}
\end{equation}
where $i^* = \mbox{argmin}_{j} \, \tilde{p}_{ji}$.
Finally, we remark that the total expenses of all agents at node $i$ in the discrete choice model are given by 
\begin{equation}
    K_i = \sum_{j=1}^N \tilde{p}_{ji} S_{ji} \, .
    \label{eq:totalex}
\end{equation}

\subsection{Transportation costs and economies of scale}

An important goal of our work is to study the role of transportation costs and scale effects on the formation of economic networks. We incorporate these aspects in terms of the consumer prices $\tilde{p}_{ji}$ that an agent at node $i$ has to pay when buying goods from node $j$.
First, we assume that consumers have to pay for the transportation of the good from the production site. As a first order approximation, transportation costs increase linearly with the distance $E_{ij} = E_{ji}$ of the nodes $i$ and $j$. Hence, the price per good for a consumer at node $i$ is the sum of the local price $\tilde{p}_{jj}$ at the production location and the transportation costs per unit good,
\begin{equation}
    \tilde{p}_{ji} = \tilde{p}_{jj} +  \tilde{p}^T E_{ji} \, . 
    \label{eq:transportation-price}
\end{equation}
The symbol $\tilde p^T$ denotes the transportation costs per unit of goods and per unit of distance. This quantity typically decreases over time as the technology in the transportation sector advances. The transportation network and the distance $E_{ji}$ are discussed in detail below. 

Second, we assume that the production is subject to economies of scale.  The higher the total production, the lower the production costs per unit. Denoting the total production at the node $j$ as $S_j$, we thus assume that $\tilde{p}_{jj}$ (the price without transportation) decreases monotonically with $S_j$. In this article, we assume an affine linear relation for the sake of simplicity
\begin{equation}
    \tilde{p}_{jj}(S_j) =  \tilde b_j - \tilde a  S_j  \, ,
\end{equation}
where the parameter $\tilde a$ describes the strength of the scale effects. We assume that the parameters $\tilde  b_j$ and $\tilde a$ are such that $\tilde{p}_{jj}(S_j) > 0$ is always satisfied.

To close the model, we express the production $S_j$ in terms of the purchases $S_{ji}$. Assuming that the production is sold completely, it must equal the purchases from all other nodes in the network such that
\begin{equation}
    S_j = \sum_i^N S_{ji} \, .
\end{equation}
For the further analysis, we define the total production at the largest supplier in the system
\begin{equation}
    S^* = \max \{S_1,\, S_2, \, ...,\, S_N\}
\end{equation}
to characterize the degree of centralization of production. 

To summarize, the consumer prices that an agent at node $i$ has to pay for goods from node $j$ are given by
\begin{equation}
    \tilde{p}_{ji}(S_j) = \left( \tilde b_j - \tilde a S_j + \tilde p^T E_{ij} \right) \, . 
    \label{eq:peff}
\end{equation}
The economic model is now complete up to the definition of the $E_{ji}$ which will be provided in the following section. Notably, the choices of the agents at a node $i$ depend on the choices of all other agents via the total production $S_j$ entering the prices. This interdependency introduces strong collective effects and essentially complicates the numerical solution as discussed in section \ref{sec:numerics}. 

In summary, we have introduced a model that describes different preferences of the agents as well as a complex cost function featuring production costs, transportation costs and economies of scale. 
The model includes three global parameters: (i) the parameter $\TT$ that describes the importance of the diversity of the agents' preferences compared to the price differences, (ii) the specific transportation costs $\tilde p^T$ and (iii) the strength of economic scale effects $\tilde a$. 

\begin{figure}[tb]
\centering
    \includegraphics[width=.65\columnwidth,clip,trim =0.cm 2.cm 0.cm 6cm]{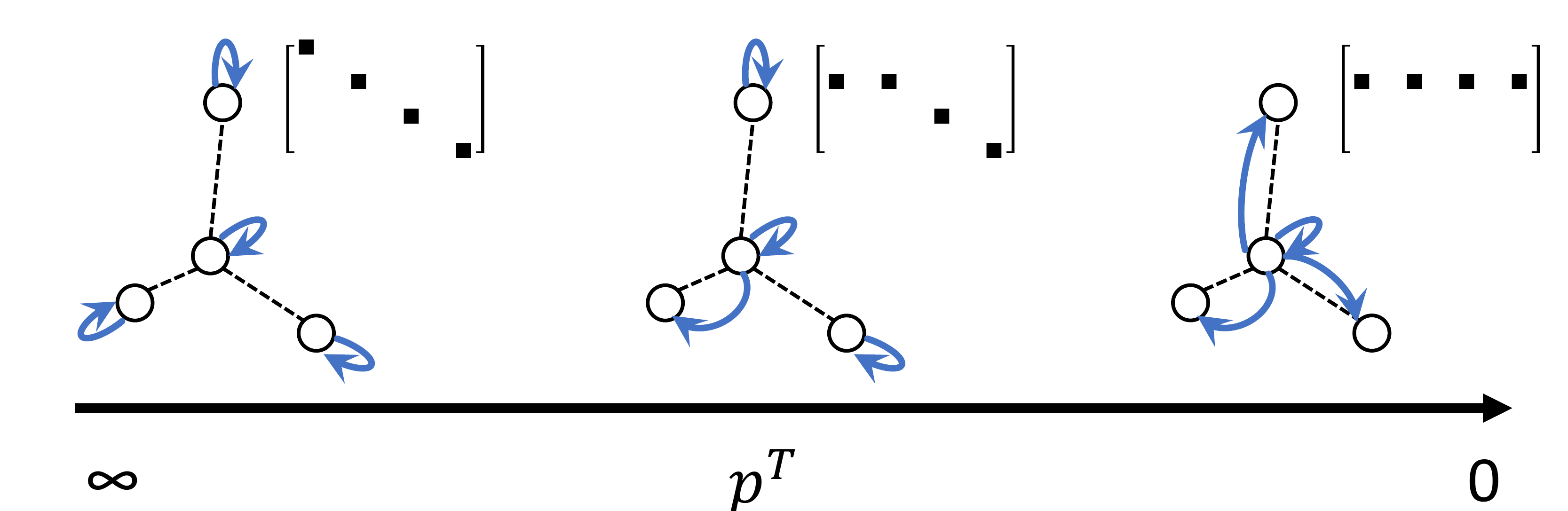}
    \includegraphics[width=.65\columnwidth,clip,trim =0.cm 2.cm 0.cm 5.5cm]{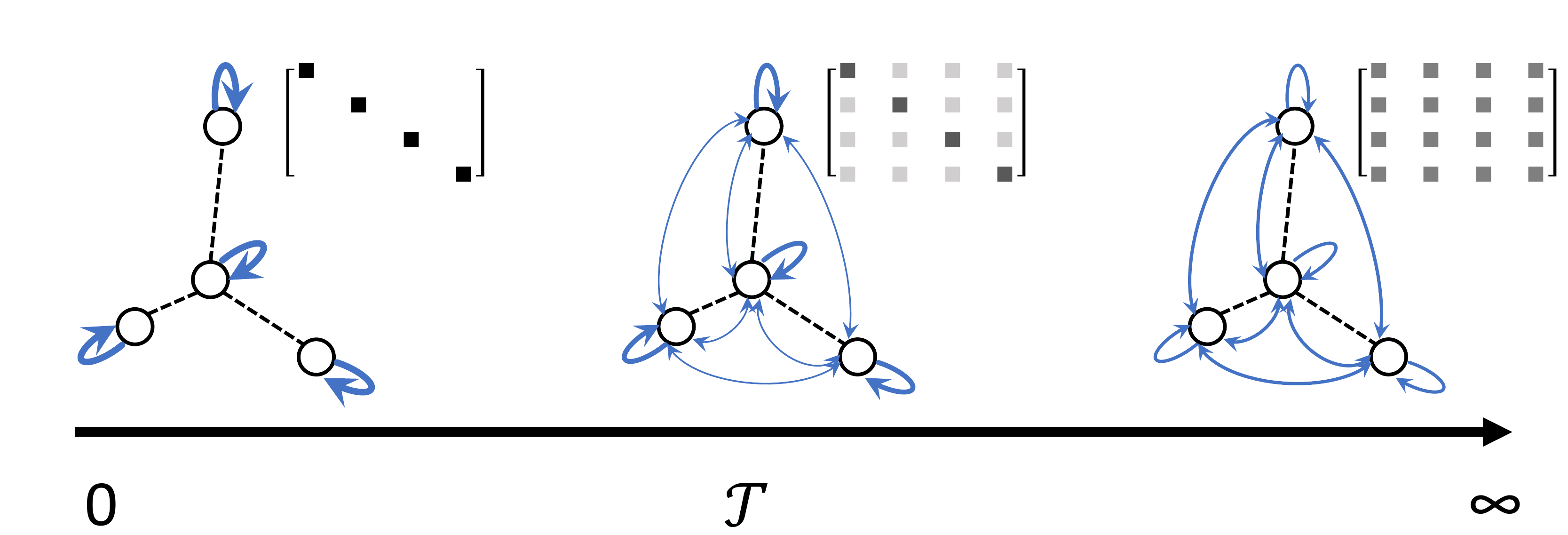}
    \caption{
    Routes to the emergence of a trade network.
    Upper row: If the specific transportation costs $\tilde p^T$ decrease, it becomes cheaper to import goods from neighboring nodes. Economic scale effects foster centralization of production and eventually lead to the emergence of a one-to-all trade network.
    Bottom row: If the diversity in preferences quantified by the parameter  $\TT$ increases, agents are choosing a diverse set of goods from other nodes despite the additional additional transportation costs which eventually leads to the emergence of an all-to-all trade network.
    The figures illustrate the existing transportation network (black) as well as the emerging trade network (blue), while the insets depict the resulting matrix of purchases $S_{ji}$.
    \label{fig:routes-to-trade}
    }
\end{figure}

The roles of the system parameters $\tilde p^T$ and $\TT{}$ are sketched in figure~\ref{fig:routes-to-trade}. 
If the specific transportation costs $\tilde p^T$ are high and the diversity of preferences $\TT$ is low, we find local production, that is $S_{ii} = D_i$ and $S_{ji} = 0$ for $i\neq j$. 
A trade network can then emerge through two different mechanisms: (i) If $\tilde p^T$ decreases, it becomes cheaper for an agent to satisfy its demand by imports than by local production.
This route to trade is strongly promoted by the scale effects of production \cite{schroder2018hysteretic}. 
Once a node starts to export goods, production costs per unit and thus the prices $\tilde{p}_{jj}(S_k)$ decrease, facilitating further exports. 
(ii) If $\TT$ increases, the preference for a diverse supply causes agents to more evenly distribute their purchases despite additional transportation costs. 
Eventually, an all-to-all connected network of trades emerges. 
We study the different routes to the emergence of a trade network in detail in sections \ref{sec:phase-diagram} and \ref{sec:transitions}.

\subsection{Transportation and trade networks}
\label{sec:trans-network}

\begin{figure}
    \centering
    \includegraphics[width=.75\columnwidth]{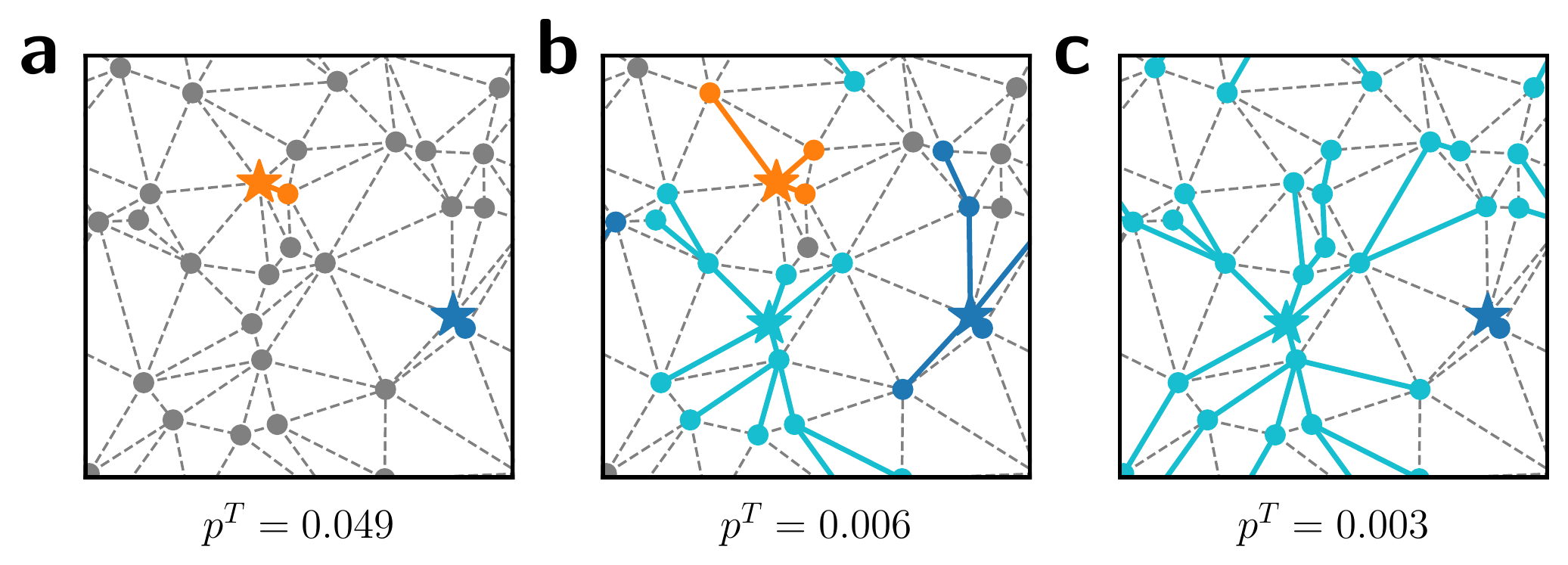}
    \caption{
    The transportation network and the emerging trade network during the centralization of production.
    Symbols and dashed lines show the nodes and the edges of the transportation network, while solid lines in different colors indicate clusters in the emerging trade network. 
    Colored stars indicate nodes that export goods to all nodes in the respective cluster, shown in the same color. 
    Grey nodes supply only themselves and gray dashed lines indicate transportation routes that are not being used.
    As the transportation cost decreases from panel \textbf{a} to \textbf{c}, the cluster size of the largest supplier grows until it encompasses almost the entire network. 
    Parameters are $\TT=0$ and $a=10^{-3}$.
    }
    \label{fig:principle_centralization}
\end{figure}

The model introduced above describes the emergence of trades in terms of the purchases $S_{ji}$ on a underlying transportation network, determining  the distances $E_{ji}$. 
To study the model numerically, we generate ensembles of geographically embedded transportation networks as follows. 
First, a number of nodes is placed uniformly at random in the unit square with periodic boundary conditions, which is equivalent to a two-dimensional torus. 
Second, these nodes are connected using Delaunay triangulation. 
The length of an edge $(i,j)$ is given by the Euclidean distance of the terminal nodes $i$ and $j$ with respect to the periodic boundary conditions. 
The distance $E_{ij}$ from node $j$ to node $i$ is then given by the geodesic distance on the transportation network.

Figure~\ref{fig:principle_centralization} shows an example of the generated transportation network as well as the emerging trade network from an exemplary simulation for $\TT=0$ and decreasing specific transportation cost $\tilde p^T$. For very high values of $\tilde p^T$ the supply matrix is diagonal, $S_{ii} = D_i$ and $S_{ji}=0$ for $j\neq i$, such that the trade network is fully disconnected. 
When $\tilde p^T$ is gradually lowered, some nodes start to purchase their goods at other nodes, represented by colored stars.
The emerging trade network is thus composed of clusters with a single supplier (identified by different colors in figure~\ref{fig:principle_centralization}). 
For very small $\tilde p^T$, the production becomes increasingly centralized such that one cluster grows until it contains all nodes.

\subsection{Aggregated Interpretation}
\label{sec:aggregated}

In this section we provide an alternative interpretation for the purchases \eqref{eq:skifull} on an aggregate level. Consider the functions
\begin{eqnarray}
    \FF_i(S_{1i},\ldots,S_{Ni}) = \FF_0 + \TT H_i(S_{1i},\ldots,S_{Ni})  - \KK_i(S_{1i},\ldots,S_{Ni}).
    \label{eq:utility}
\end{eqnarray}
for all nodes $i=1,\ldots,N$ with a constant $\FF_0$ and
\begin{eqnarray}
    H_i &= - \sum_{\ell=1}^N \frac{S_{\ell i}}{D_i}\ln{\frac{S_{\ell i}}{D_i}} , 
       \label{eq:entropypurchase} \\
    \KK_{i} &= \sum_{\ell=1}^{N} \left( \tilde b_\ell
    - \tilde a( S_\ell - S_{\ell i}/2)+ \tilde p^T E_{i \ell} \right)
    \frac{S_{\ell i} }{D_i}  .
     \label{eq:costs-total}
\end{eqnarray}
Now consider the purchases that maximize the function $\FF_i$ while respecting the constraint 
\begin{equation*}
    \sum_{\ell=1}^N S_{\ell i} = D_i ,
\end{equation*}
That is, all purchases made by agents at the node $i$ must sum to $D_i$. The maximum can be computed by using the method of Lagrangian multipliers, leading to the conditions
\begin{equation}
    \frac{\partial}{\partial S_{ji}} \left[ \FF_i - \lambda \left( \sum_{\ell=1}^N S_{\ell i} - D_i \right) \right] = 0.
\end{equation}
Solving these conditions yields
\begin{equation}
    S_{ji} = D_i \, \frac{\exp(- \tilde{p}_{ji}/\TT)} 
    {\sum_{\ell = 1}^N\exp(- \tilde{p}_{\ell i}/\TT)} .
    \label{eq:skifull2}
\end{equation}
which is equivalent to the expression \eqref{eq:skifull}. 

The maximum of the function $\FF_i$ has be to be evaluated for all nodes $i=1,\ldots,N$ -- but these optimization problems are not independent. Every function $\KK_i$ depends on the production $S_j$ at all nodes and hence on the results of the optimization problems of \emph{all} nodes in the network. We thus have to interpret the purchases  \eqref{eq:skifull2} as a Nash equilibrium: No node $i$ can further increase the function $\FF_i$ by changing its purchases $S_{1i},\ldots,S_{Ni}$ while the purchases of all other nodes remain constant. 

We have thus introduced an alternative, macroscopic approach that reproduces the aggregated purchases obtained from discrete choice theory. The aggregated purchases of node $i$ maximize the function $\FF_i(S_{1i},\ldots,S_{Ni})$, which may thus be interpreted as an aggregated effective utility function. We propose the following interpretation of this aggregated utility, proceeding term by term. 
First, there is a constant term $\FF_0$ describing the utility from using a good. Second the function $H_i$ coincides with the Gibbs entropy, which measures the diversity of the purchases. Hence, the aggregated utility increases with the diversity of supply. The strength of this effect is measured by the parameter $\TT{}$.
Finally, the utility is reduced by the function $\KK_i$. We consider the case of a large network with $N$ nodes where production is not fully centralized such that
\begin{equation}
    S_{\ell i} \ll S_{\ell} .
\end{equation}
Then we have
\begin{eqnarray}
    \KK_{i} &= \sum_{\ell=1}^{N} \left( \tilde b_\ell
    - \tilde a( S_\ell - S_{\ell i}/2)+ \tilde p^T E_{i \ell} \right)
    \frac{S_{\ell i}}{D_i}  \\
    &\approx \frac{1}{D_i} \sum_{\ell=1}^N \tilde{p}_{\ell i} S_{\ell i}.
    \label{eq:costs-approx}
\end{eqnarray}
Hence, $\KK_i$ is approximately equal to the total expenses \eqref{eq:totalex} of the agents at node $i$ divided by the number of agents $D_i$.

We finally note that the negative function, $-\FF_i =  \KK_i  - \TT H_i$, has a similar form as the Helmholtz free energy in the study of closed thermodynamic systems. 
Because of this structural similarity, we refer to the weighting factor $\TT$ as the effective temperature of the economic system in the following. 
The similarities to statistical physics will further guide our analysis of the system and provide methods to quantitatively understand the transitions between different trade regimes.

\subsection{Numerical solution}
\label{sec:numerics}

The equilibrium state of the trade network will be analyzed via numerical simulations. In all numerical simulations we fix the system parameters as follows. First, we assume that the nodes are chosen such that they contain the same number of agents such that 
\begin{equation}
    D_i = D = \frac{D_{\rm tot}}{N},
\end{equation}
where $D_{\rm tot}$ is the total number of agents. Furthermore, we will a denote the purchases $S_{ji}$ and productions $S_i$ in units of $D_{\rm tot}$. In these rescaled units, all purchases sum up to one, $\sum_{j,i} S_{ji} = 1$. Furthermore, we have $S_{ii}=1/N$ when the purchases are fully local (i.e. $S_{ji}$ is diagonal) and $S^* = 1$ when the production is fully centralized at a single node. 

The prices $\tilde p_{ji}$ depend on three essential parameters, $\tilde a$, $\tilde b_j$ and $\tilde p^T$. For consistency with prior work, we scale all these parameters with the factor $D$,
\begin{equation}
    \tilde p_{ji} = D p_{ji}, \quad
    \tilde a = D a, \quad
    \tilde b_j = D b_j, \quad
    \tilde p^T = D p^T.
\end{equation}
Hence, the approximation \eqref{eq:costs-approx} now reads $\KK_i = \sum_\ell p_{\ell i} S_{\ell i}$. The parameter $b_i$ is chosen uniformly at random from an interval $b_0 + [0, 0.005]$ for each node. The parameters $a$ and $p^T$ are varied to analyze how scale effects and transportation costs scale the emerging trade network. In the simulations, we use $N=300$ unless stated otherwise.

In the model developed in the previous section, purchases and prices are linked via the conditions \eqref{eq:peff} and \eqref{eq:skifull}. Notably, both equations are coupled: The purchases depend on the prices, but the prices also depend on the purchases through the production $S_j$. Both equations have to be solved self-consistently.

Based on these considerations, we establish the following self-consistent algorithm to compute the equilibrium purchases as a function of the system parameters. 
Starting from a suitable initial guess for the purchases $S_{ji}$, we compute the resulting prices \eqref{eq:peff}. 
Given these prices we can directly compute a new value for the purchases \eqref{eq:skifull}. 
This procedure is iterated until no further changes in the purchases occur. 
Once the iteration converged, the resulting state satisfies both conditions \eqref{eq:peff} and \eqref{eq:skifull} simultaneously such that we arrived at an equilibrium.

We emphasize that the model can support multiple solutions~\cite{schroder2018hysteretic}. 
In such a situation it depends on the initial guess which one is found and if the iteration terminates at all. 
In the numerical simulations, we use the following algorithm to compute how the equilibrium states depends on the parameters $a$, $p^T$ and $\TT$:
\begin{enumerate}
\item We fix a value of $a > 0$ and a transportation network as described above.

\item We start at $\TT = 0$ and $p^T = \infty$, where the equilibrium state is given by $S_{ii} = 1/N$ and $S_{ji}=0$ for $j\neq i$, i.e.~fully local production.

\item \label{alg:old_result} We then compute the equilibrium states along the line $\TT = 0$ by decreasing $p^T$ to zero. 
This computation follows the semi-analytic algorithm introduced in \cite{schroder2018hysteretic}.

\item We define a grid of values for $p^T$ and $\TT$ for which the supply matrix $S_{ji}$ is to be computed. 
We choose the minimum (maximum) value of $p^T$ such that production is fully centralized (local) for $\TT = 0$. 
The step size is chosen uniformly on a logarithmic scale. 
We then compute the equilibrium states as a function of $\TT$ and $p^T$:

\begin{enumerate}
    \item For each value of $p^T$ on the grid, we proceed from $\TT = 0$ to $\TT = \TT_{\rm max}$. 
    \item For a given value of $p^T$ and $\TT$ we start from an initial guess for the purchases $S_{ji}$. 
    For $\TT>0$ we use the solution of the previous, smaller value of $\TT$. For $\TT=0$ the exact solution $S_{ji}$ is known from the step (\ref{alg:old_result}). 
    \item We compute the prices $p_{ji}$ from equation~\eqref{eq:peff} and update the purchases $S_{ji}$ using equation~\eqref{eq:skifull}.
    \item We iterate this procedure until it converges.  
    Convergence is assumed when the Frobenius norm of the difference of the previously calculated and the updated purchase matrix decreases below $10^{-8}$.
\end{enumerate}
\item The procedure is repeated for different random realizations of networks to average out the impact a single network might have on the position and behavior at the phase transition. 

\end{enumerate}

\section{Phase diagram of the trade network}
\label{sec:phase-diagram}

How do the decisions of individual agents lead to the emergence of trade? 
In this section we provide an overview of the emerging trade networks and how they depend on the preferences of the agents, the properties of the transportation network, and the economic scale effects. 
To this end, we compute the equilibrium purchases $S_{ji}$ using the self-consistent method introduced in section \ref{sec:numerics} as a function of the parameters $p^T$, $\mathcal{T}$, and $a$. 
We average over $100$ random realizations of the transportation network model (cf.~section \ref{sec:trans-network}) to map out the typical behavior independent of the randomly chosen network. 
Single realizations will be treated in a subsequent section. 

\begin{figure*}[tb]
    \centering
    \includegraphics[trim = 0cm 1.5cm 0cm 1cm, width=0.9\textwidth]{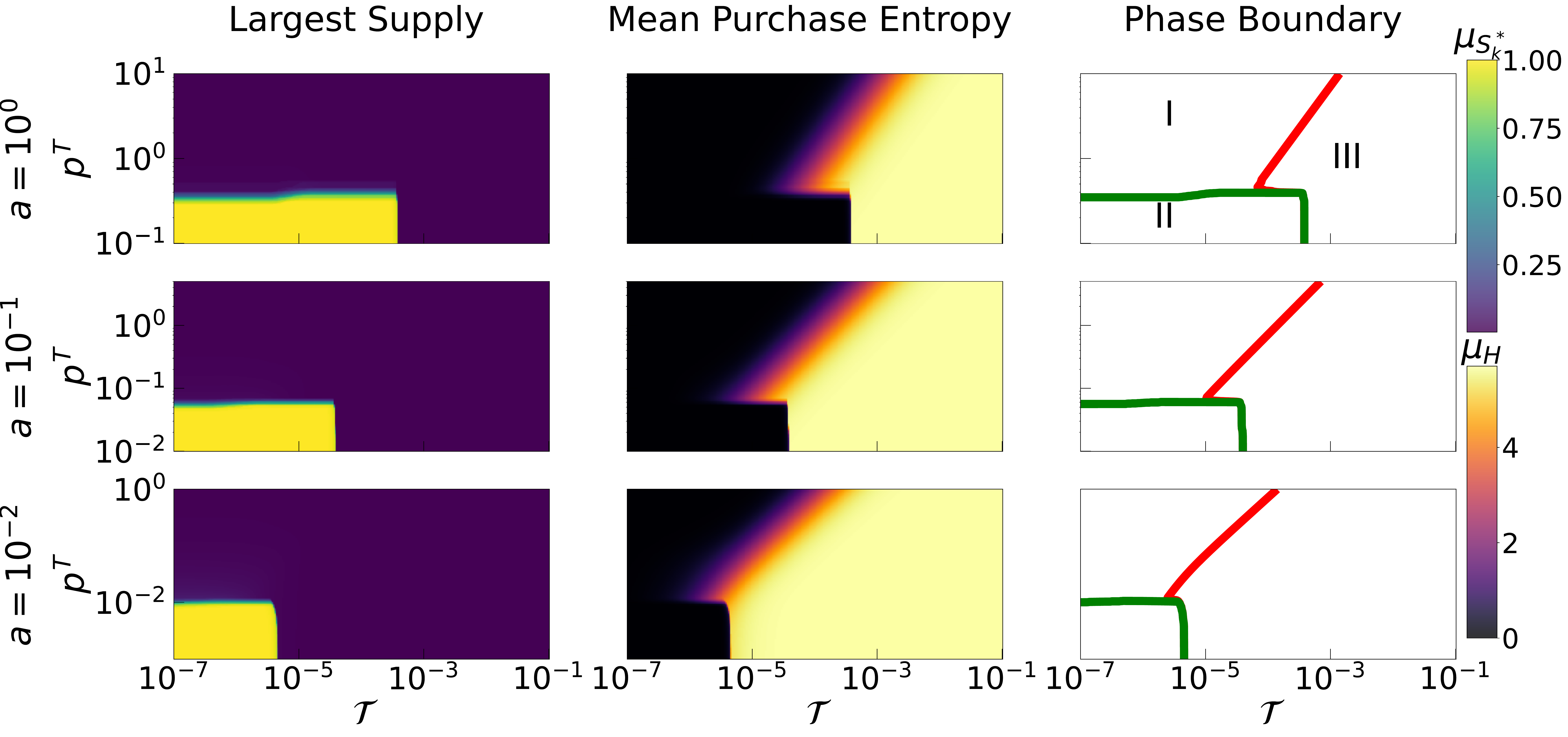}
    \caption{Phase diagrams of the emerging trade networks for different values of the scale effects $a$.  
    The left column shows the production of the largest supplier in the system $\mu_S$, and the middle column shows the mean purchase entropy $\mu_H$. 
    Results for $\mu_S$ and $\mu_H$ are averaged over $100$ random realizations of the transportation network.
    Based on these observations, we identify three phases of the emerging trade network shown in the right column:
    (I) A phase of local production when the specific transportation costs $p^T$ are high and the effective temperature $\TT$ is low.
    (II) A phase with centralized production, i.e.~$\mu_S \rightarrow 1$, for small values of both $p^T$ and $\mathcal{T}$. 
    (III) A phase with diversified production, i.e.~high entropy for large values of $\mathcal{T}$, i.e. $\lim_{\mathcal{T} \rightarrow \infty} \, \mu_H = \ln(N)$.
    A definition and analysis of the phase transition is provided in the main text.
    }
    \label{fig:phase_diagrams}
\end{figure*}

Figure~\ref{fig:phase_diagrams} shows two macroscopic observables that characterize the emerging trade network on large scales. 
The normalized maximum production, 
\begin{equation}
    \mu_S =  \left\langle S^* \right\rangle,
\end{equation}
quantifies the degree of centralization of the production, where the brackets denote the average over the transportation network ensemble. 
If $\mu_S$ is close to unity, the production is strongly centralized at a single node. 
The average entropy
\begin{equation}
    \mu_H = \left\langle N^{-1} \sum_{i=1}^N H_{i} \right\rangle
\end{equation}
measures the diversity of supply as described in section \ref{sec:trade-model}. 

Based on the observables $\mu_S$ and $\mu_H$, we identify three different regions in parameter space that display qualitatively different patterns of the emerging trade networks:
\begin{enumerate}
    \item[I.] If the specific transportation costs $p^T$ are high and the effective temperature $\TT$ is low, we have
    \begin{eqnarray}
        & \lim_{\mathcal{T} \rightarrow 0}\,\lim_{p^T \rightarrow \infty} \, \mu_S = \frac{1}{N} \quad \mbox{and} 
        \nonumber \\ 
        & \lim_{\mathcal{T} \rightarrow 0}\,\lim_{p^T \rightarrow \infty} \, \mu_H = 0.
    \end{eqnarray}
    That is, all agents purchase their goods locally as imports would be too expensive. No trading takes place in this phase such that the purchase matrix is given by
    \begin{equation}
        \lim_{\mathcal{T} \rightarrow 0}\,\lim_{p^T \rightarrow \infty} \, 
        S = \frac{1}{N} \left( \begin{array}{c c c c}
        1 & 0 & \dots & 0 \\
        0 & 1 & \dots & 0 \\
        \vdots & \vdots & \ddots & \vdots \\
        0 & 0 & \dots & 1
        \end{array} \right).
        \label{eq:Smat-local}
    \end{equation}
    We refer to this phase as the \emph{local production phase} in the following.
    \item[II.] If the unit transportation costs $p^T$ decrease, while the effective temperature $\TT$ is still low, we find 
    \begin{equation}
        \lim_{\mathcal{T} \rightarrow 0}\,\lim_{p^T \rightarrow 0} \, \mu_S = 1.
    \end{equation}
    That is, the production becomes completely centralized at a single node $j$ and the purchase matrix can be written as
    \begin{equation}
        \lim_{\mathcal{T} \rightarrow 0}\,\lim_{p^T \rightarrow 0} \, S = \frac{1}{N}  
        \left( \begin{array}{c c c c c}
          0 & \dots & 0 & \dots & 0 \\
          1 & \dots & 1 & \dots & 1 \\
            \vdots & \vdots & \vdots & \vdots & \vdots \\
            0 & \dots & 0 & \dots & 0
        \end{array} \right).
        \label{eq:Smat-central}
    \end{equation}
    We refer to this phase as the \emph{centralized production phase} in the following.
    \item[III.]
    If the effective temperature $\TT$ is high, the differences of the agents' preferences dominate while different prices play a negligible role. Hence, we observe a phase with an average entropy close to the maximum possible value,
    \begin{equation}
        \lim_{\mathcal{T} \rightarrow \infty} \, \mu_H = \ln(N).
    \end{equation}
    In this phase every node purchases a similar amount of goods from every other node such that the purchase matrix reads
    \begin{equation}
        \lim_{\mathcal{T} \rightarrow \infty} \, S = \frac{1}{N^2} 
        \left( \begin{array}{c c c c c}
            1 & \dots & 1 & \dots & 1 \\
            1 & \dots & 1 & \dots & 1 \\
            \vdots & \vdots & \vdots & \vdots & \vdots \\
             1 & \dots & 1 & \dots & 1
        \end{array} \right).
        \label{eq:Smat-diverse}
    \end{equation}
    We thus find a globally connected trade network, where every node produces the same amount of goods, $\lim_{\mathcal{T} \rightarrow \infty} \, \mu_S = 1/N$, as in phase I but exports and imports from all other nodes. We refer to this phase as the \emph{diverse production phase} in the following.
\end{enumerate}

For a better overview, we extract a comprehensive phase diagram from the values of $\mu_S$ and $\mu_H$ in our simulations as follows. 
For both quantities, we determine the minimum and maximum values found in the simulations and choose the midpoints of the respective intervals as a threshold value. 
We then compute the areas in the phase diagram, where the observables $\mu_S$ and $\mu_H$ are above or below the respective thresholds. 
The red and green lines in figure~\ref{fig:phase_diagrams} depict the boundary between these areas. 
Together, they reveal the boundary between the different phases I, II and III. 

The resulting phase diagram in figure~\ref{fig:phase_diagrams} on the right shows most clearly how the equilibrium trade network depends on the parameters $p^T$ and $\TT$ as well as the strength of the scale effects $a$. 
We find that the three phases exist for all values $a$, but the location and nature of the phase boundaries depend strongly on $a$. 
Scale effects foster a centralization of production, such that the parameter region corresponding to centralized production (phase II) increases with $a$.

We observe qualitatively different transitions between the three phases:
\begin{itemize}
\item
The transition between local and diverse production (I-III) is smooth. The phase boundary is a straight line $p^T / \TT = \mathrm{const}$.
\item
The transition from local to centralized production (I-II) is rather sharp and the phase boundary is almost given by a horizontal line, i.e. a constant value of $p^T$. 
However, a slight incline of the phase boundary becomes visible for large $a$, such that the transition can in principle also be triggered by a decrease in $\TT$.

In economic terms, centralization of production is driven by a reduction in the specific transportation costs $p^T$. 
This process is facilitated by scale effects, which make the transition occur earlier (i.e.~for larger values of $p^T$) and more rapid. 
This scenario is mostly, but not perfectly, independent of the consumer preferences expressed by the parameter $\TT$.
\item
The transition between centralized and diverse production (II-III) is also very sharp and the phase boundary is almost given by a vertical line, i.e. a constant value of $\TT$. 

In economic terms, a change in the nodes' preferences can trigger a transition at low specific transportation costs $p^T$. If the agents' preferences vary only little (low $\TT$), then decisions are dominated by prices and production is centralized at a single node. 
If the agents' preferences vary strongly (high $\TT$), prices play a minor role and the production is decentralized. 
Remarkably, the transition occurs abruptly as $\TT$ increases.

\item 
A particular behavior is observed around the triple point of the phases I, II, and III. For certain values of $\TT$, we find a non-monotonic behavior of $\mu_H$. 
Decreasing the unit transportation costs first induces a transition from local to diverse production, and then a transition to centralized production. 
Hence, $\mu_H$ first increases and then decreases back to values around zero. 
We note, however, that the system can have multiple equilibria \cite{schroder2018hysteretic}, of which only one is analyzed here. 
\end{itemize}

\section{Transitions between different regimes}
\label{sec:transitions}

We now investigate the transitions between the three phases of the trade network in more detail. We discuss the type of the transition and derive approximate analytic expressions for the locations of the phase boundaries.

\subsection{From local to centralized production}
\label{sec:transition-I-III}

The transition from phase I to phase II describes the centralization of production as the price for transportation decreases, while the diversity of the agents' preferences does not play a central role. 
The remarkable feature of this transition is that it can be either continuous or discontinuous depending on the value of $a$ \cite{schroder2018hysteretic}. 
A discontinuity is the direct consequence of scale effects in the production. 
If agents at a node $i$ chooses to purchase goods at a node $j$ instead of locally, the production costs per unit at node $j$ decrease due to scale effects. Now the agents at another node $i'$ can purchase goods from $j$ at a lower price. If this effect is strong enough, agents at $i'$ may also choose to change their purchases and buy at $j$ instead, which leads to a further decrease of production costs at $j$. Eventually, we may find a cascade of decisions, where a large fraction of agents simultaneously change their purchases and the production is centralized at node $j$. 

We analyze the transition in more detail, starting from the simplest case $\TT=0$ \cite{schroder2018hysteretic}. In this case the differences between the preferences of agents vanish and all agents at a node $i$ purchase from same node $i^*$ which yields the lowest price, see equation~\eqref{eq:Sji-Tzero}. We will thus treat a node as a single entity in the following analysis. 

If $a=0$ (and $\TT=0$) the maximum production $S^*$ changes in discrete steps of $D$ as $p^T$ is reduced. For $\TT=0$, all agents at a node $i$ have the same preferences and choose a single supplier node. If node $i$ chooses to buy at node $j$ instead of a node $\ell$, the effective price node $i$ pays per units changes by
\begin{equation}
    \Delta p = p_j - p_\ell =
    (b_j-b_\ell) + p^T (E_{i j} - E_{i \ell}).
    \label{eq:change-peff-il}
\end{equation}  
Hence, a node $i$ will change its purchases if $p^T = (b_j-b_\ell)/(E_{ij} - E_{i \ell})$. 
These values of $p^T$ are distinct for all nodes and potential suppliers with probability 1, such that changes in the purchases occur only in single events at different values of $p^T$. 
Hence, in the thermodynamic limit $N \rightarrow \infty$  with constant total demand $ND$ constant, the transition from phase I to phase II is continuous \cite{schroder2018hysteretic}.

If $a$ is large (and $\TT=0$), the transition is generally discontinuous in the following sense. 
If $p^T$ is reduced, the maximum production $S^*$ changes by a macroscopically large amount $\Delta S^*$ at a critical value of $p^T$. 
This is due to a cascade of decisions of the individual nodes: If a node $i$ chooses to purchase from another node $j$ instead of buying locally, the price $p_{jj}$ at node $j$ decrease by an amount $aD$. 
This decrease may be sufficiently strong to cause another node $i'$ to also purchase from $j$ instead of buying locally. 
In the end, a macroscopic fraction of the nodes decides to change its purchases simultaneously. 
We note that the difference between continuous or discontinuous transitions is not visible in figure~\ref{fig:phase_diagrams}, as the phase diagram shows the average over many random realizations of the underlying transportation network. 
In contrast, a clear difference is observed for a single realization as shown in figure~\ref{fig:transition-I-II-single}.

For extremely large values of $a$ we may even find a complete centralization in a single event, i.e.,~$\Delta S^*=(N-1)D$. 
We make this statement more precise now. 
To this end, we order the nodes $n\in\{1,\ldots,N\}$ as follows. 
The nodes $(j,i) = (1,2)$ are chosen such that they have the lowest value of $(b_j-b_i) + E_{ji}$, that is
\begin{equation}
    (b_j-b_i) + E_{ji} \le (b_m-b_n)  + E_{mn}, 
    \quad \forall \, m \neq n.
    \label{eq:centralization-first-pair}
\end{equation}
The remaining nodes $n\in\{3,\ldots,N\}$ are ordered such that the series
\begin{equation}
    (b_1 - b_n) + E_{1n}
\end{equation}
is monotonically increasing. 
Furthermore, we assume that 
\begin{equation}
    (i - 1) E_{12} > E_{1i}, \qquad \forall \, i \in \{3,\ldots,N \}.
\end{equation}
This condition is typically satisfied if the differences in the parameters $b_i$ are small.
Then we find the following statement: If scale effects are extremely strong, 
\begin{equation}
   a > \max_{i\in\{3,\ldots,N\}} \frac{|(b_1-b_i) E_{12} - (b_1-b_2) E_{1i}|}{D \left[ (i-1)E_{12} - E_{1i} \right]} \, ,
   \label{eq:centralization-discontinuous-a}
\end{equation}
the maximum production changes from $S^* = D$ to $S^*=ND$ when the transportation costs per unit are decreased below a critical value
\begin{equation}
    p^T_{\rm crit} = \frac{aD}{E_{12}} \,.  
    \label{eq:ptcrit-I-II-a-large}
\end{equation}
A proof of this statement is given in~\ref{sec:proof-disc}. For large $a$ we may thus compute the critical point directly from the network topology and the local values $b_i$.

\subsection{Impact of temperature on the centralization}
\label{sec:transition-I-III-T}

\begin{figure}[tb]
    \centering
    \includegraphics[trim=0cm 5cm 0cm 0cm, width=.75\columnwidth]{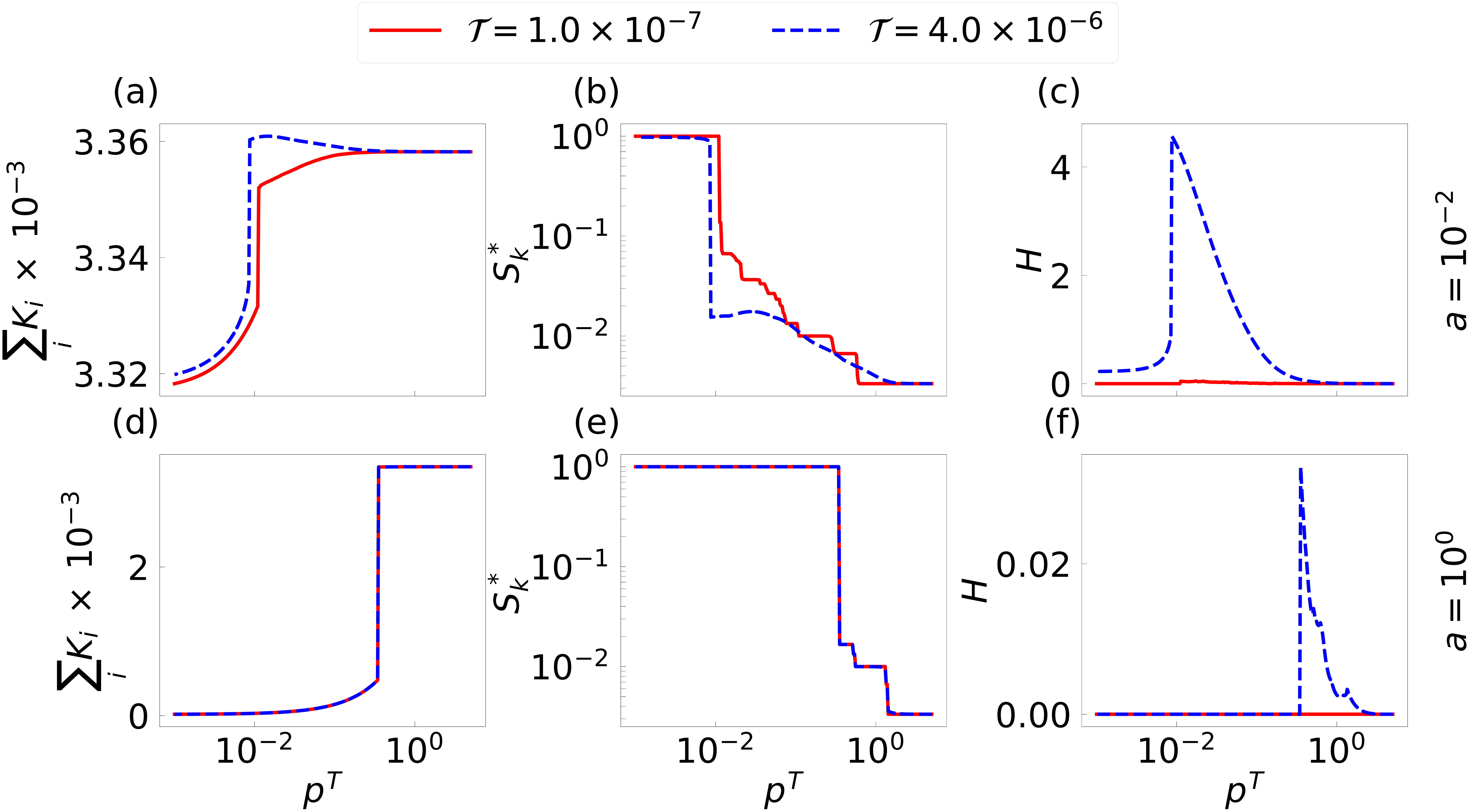}
    \caption{
    Transition from local to centralized production. 
    Results are shown for a single instance of the transportation network, comparing the case of (a-c) weak scale effects and (d-e) strong scale effects as well as low and high effective temperatures (red solid vs.~blue dashed line).
    We show (a,d) the total expenses $\sum_i K_i$, (b,e) the production of the busiest node $S^* = \max_k S_k$ and (c,f) the entropy  averaged over all nodes, $H= N^{-1} \sum_i H_i$.
    \label{fig:transition-I-II-single}
    }
\end{figure}

A reduction in the specific transportation costs $p^T$ generally leads to a centralization of production. 
But how does the diversity of agent's preferences, measured by the parameter $\TT$, affect this scenario? 
Simulation results for different values of $\TT$ and $a$ are shown in figure~\ref{fig:transition-I-II-single}.

In the case of strong scale effects, $a=10^0$, we observe almost no differences between the curves for different $\TT$. 
In this case, the price effects dominate the decision of the agents: The centralization of production leads to strong changes of the price $p_{ji}$ which outweighs the individual differences $\TT \epsilon_j$ in the agents' utility function \eqref{eq:utility-agent}.

In contrast, a substantial impact is found in the case of weak scale effects, e.g. $a=10^{-2}$. Results for two intermediate values for $\TT{}$ are compared in the upper row of figure~\ref{fig:transition-I-II-single}.
Not surprisingly, the differences in the agents' preferences measured by $\TT \epsilon_j$ counteracts centralization, as the impact of price differences due to scale effects is less pronounced. 
Hence, the maximum production $S^*$ for intermediate values of $p^T$ is typically the lower, the higher $\TT{}$. Correspondingly, there are more nodes that contribute significantly to the production. 
Nevertheless, we still see a transition to complete centralization for the given values of $\TT{}$ as $p^T$ decreases further. Remarkably, the final step of the centralization process takes place in an even more abrupt way as $\TT{}$ is increased. 
The difference is even more pronounced in the aggregated entropy $H = \sum_i H_i$. For $\TT{} = 4 \times 10^{-6}$ we first observe a smooth increase of $H$, indicating the emergence of all-to-all trade, until it drops sharply to low values associated with centralization.
We thus find that the diversity of preferences can delay centralization of the trade network, until the centralization occurs in an `explosive' way. 
Similar explosive effects where found for a variety of different models in percolation theory \cite{achlioptas2009explosive,souza2015anomalous}.

We finally recall that for even higher values of $\TT{}$ the emergence of centralization is entirely absent as shown in the phase diagrams in figure~\ref{fig:phase_diagrams}.

\subsection{From local to diverse production}

The differences in the agent's preferences induces a transition from local production to a diverse supply if either the effective temperature $\TT$ is increased or the specific transportation costs $p^T$ are decreased. The boundary between the two regimes appears as a straight line with slope one in the double-logarithmic phase diagram (figure~\ref{fig:phase_diagrams}), except for the parameter region with both $p^T$ and $\TT$ small leading to centralized production (phase II). 
In the following paragraphs, we provide a detailed analysis of the transition from local to diverse production and derive an analytical estimate for the location of the transition.

We first note that in both phases the local production of each node equals $S_j = D$. 
Hence, we assume that scale effects play no role for the transition and equation~\eqref{eq:skifull} for the purchases simplifies to
\begin{equation}
    S_{ki} = D \, \frac{ \exp\left[ -D (b_k + p^T E_{ki}) / \TT \right]  }{
    \sum_{k=1}^N  \exp\left[ -D (b_k + p^T E_{ki}) / \TT \right]  }.
    \label{eq:Ski-without-scale}
\end{equation}
The differences in the local parameters $b_k$ are small compared to $p^T E_{ki}$ in all simulations and can thus be neglected in the following analysis. 
We conclude that the transportation costs parameter $p^T$ and the effective temperature $\mathcal{T}$ enter only via the ratio 
\begin{equation}
    \beta = D \, p^T / \TT \, .
\end{equation}

Hence, also the entropy $H$ will depend on $p^T$ and $\TT$ only via the ratio $\beta$ and the phase boundary is given by a straight line
\begin{equation}
    p^T / \TT = D^{-1} \, \, \beta_{\rm crit} =  {\rm const.}
\end{equation} 
Indeed, the numerical simulations presented in figure~\ref{fig:phase_diagrams} confirm this finding. 
The boundary between phases I and III is a straight line as long as $p^T$ and $\TT$ are large enough such that no centralization occurs, i.e., far away from phase II.

The second conclusion we draw from the expression \eqref{eq:Ski-without-scale} is that we can treat all nodes $i=1,\ldots,N$ separately. 
We note that this assumption is strictly true only for $a=0$, while it represents a useful approximation for $a > 0$. 
Indeed, the success of this assumption is surprising from a conceptual view at first glance. 
If a node $i$ would independently redistribute its purchases, this would invalidate the assumption $S_j \approx D = \mathrm{const.}$ which allowed us to neglect scale effects and treat all nodes separately. 
This apparent contradiction is resolved as follows: Typically, The decision of a node $i$ to purchase from $j$ is mirrored by a simultaneous decision of $j$ to purchase at $i$ if $b_i$ and $b_j$ do not differ too much. 
Hence, the decisions are not independent a priori but their effect on the prices cancels out such that the decisions of the nodes effectively separate and can be treated independently in the calculation.

Using these assumptions, we now provide an explicit approximate expression for the entropy $H_i$ and the critical parameter $\beta_{\rm crit}$. 
We first note that the entropy can be rewritten as (cf.~\ref{sec:app-entropy})
\begin{equation}
    H_i = - \frac{\partial}{\partial \beta^{-1}} \big[ - \beta^{-1} \ln(Q_i) \big],
    \label{eq:H-from-Q}
\end{equation}
where 
\begin{equation}
    Q_i = \sum_{j =1}^N e^{-\beta E_{ji}}
    \label{eq:partition}
\end{equation}
can be interpreted as a partition function and the expression in the bracket as a free energy. 
To evaluate the $Q_i$, we just need the information of how many nodes are found at a given distance $E$. 
We encode this in the counting function 
\begin{equation}
    \mathcal{N}_i(\hat E) = \sum_{j =1}^N \Theta(\hat E - E_{ij}),
    \label{eq:counting-fun}
\end{equation}
where $\Theta$ denotes the Heaviside function. 
Notably, the function \eqref{eq:counting-fun} can be interpreted as an integrated density of states. 

\begin{figure}
    \centering
    \includegraphics[trim = 0cm 0cm 0cm 0cm, width=.75\columnwidth]{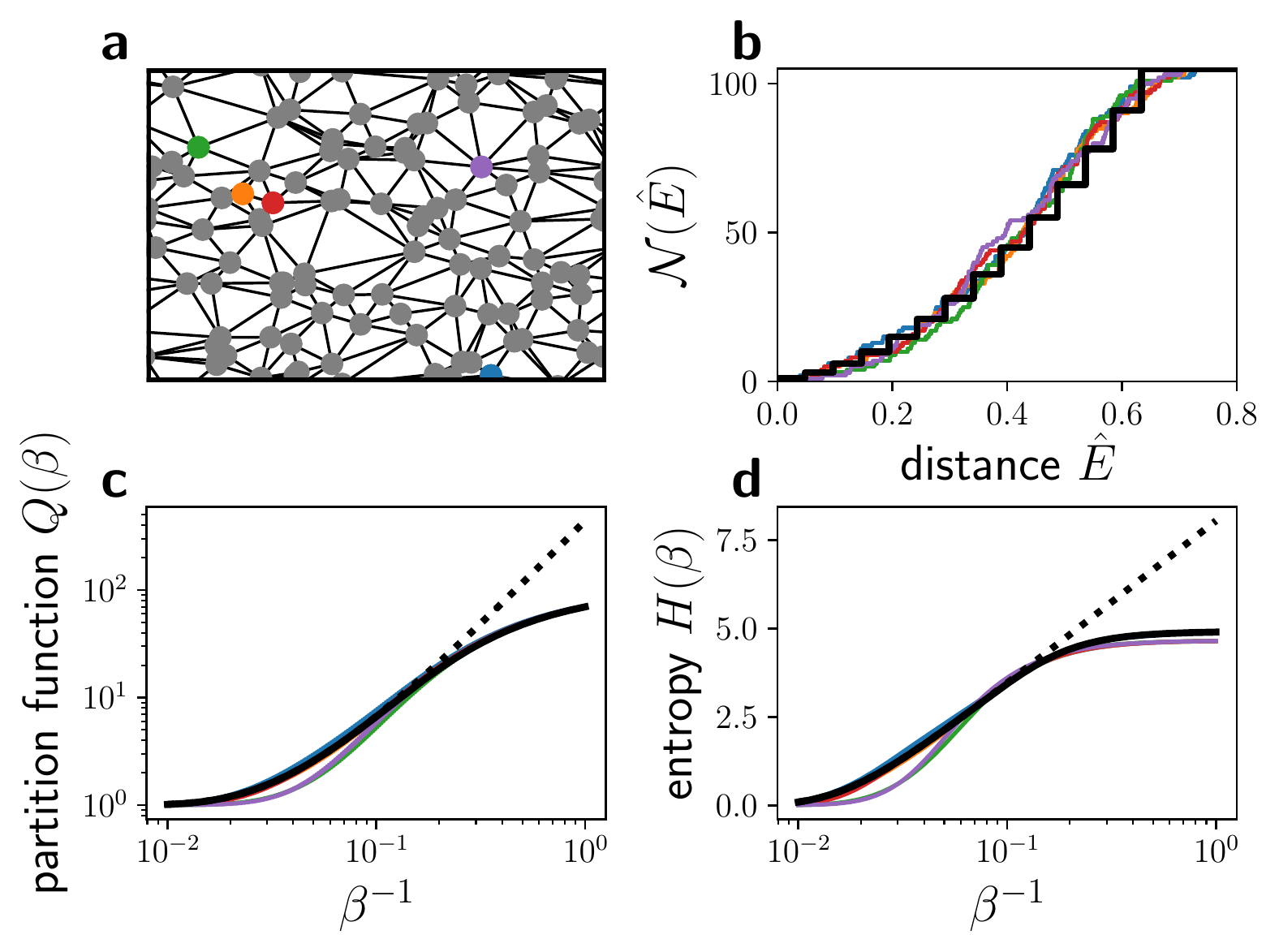}
    \caption{
    Transition between local and distributed production in a sample network.
    (a) Network structure of the sample network with $N=105$ nodes. Five nodes were selected at random for further analysis (color coded).    
    (b) The counting function $\mathcal{N}(\hat E)$ counts the number of nodes with a distance $\le \hat E$. We show $\mathcal{N}_i(\hat E)$ for the five selected nodes (thin colored lines) in comparison to the analytic approximation \eqref{eq:N-staircase} (thick black line).
    (c,d) The partition function \eqref{eq:partition} and the entropy \eqref{eq:H-from-Q} for the five selected nodes (thin colored lines) compared to the analytic approximation \eqref{eq:Q} (thick black line).
    On average, we find a very good agreement. 
    For low temperatures, the partition function and entropy can be approximated by the expressions \eqref{eq:Q-high-beta} and \eqref{eq:H-high-beta} (dotted black line).
    }
    \label{fig:transition-entropy}
\end{figure}

To obtain an analytic approximation for $Q_i$, we have to approximate $\mathcal{N}_i$ by a function that keeps the essential properties but allows to carry out the sum in equation \eqref{eq:partition} in closed form.
Three properties have to be taken into account: (i) The function $\mathcal{N}_i$ scales quadratically with $\hat E$ on coarse scales.
(ii) We have to take into account that the set of distances $E_{ij}$ is discrete such that $\mathcal{N}_i$ does not increase smoothly but in discrete steps. 
In fact, it is essential to take into account that the distance to the nearest neighbor is always finite. 
In an analog physical model, this would correspond to a finite `energy gap' between the ground and first excited state. 
(iii) Finally, it must be taken into account that the number of states is finite. 
These requirements can be met by a staircase function with steps of regular position and size. 
For the time being, we restrict ourselves to networks where the number of nodes can be written as $N = (M+1)(M+2)/2$ with $M \in \mathbb{N}$. 
The function $\mathcal{N}_i(\hat E)$ can then be approximated by
\begin{equation}
    \mathcal{N}_{\rm st}(\hat E) = \sum_{m=0}^M (m+1) \Theta(\hat{E} - m E_0),
    \label{eq:N-staircase}
\end{equation}
where $E_0 = 1/\sqrt{4N}$ is the expected value of the distance to the nearest neighbor on a two-dimensional plane with node density $\rho=N$. 
Using this approximation, the partition function can be computed as
\begin{eqnarray}
    Q_{\rm st} &= \sum_{m=0}^M (m+1) e^{-\beta E_0 m}  \nonumber \\
    &=\frac{1- (M+2) e^{-\beta E_0 (M+1)} + (M+1) e^{-\beta E_0 (M+2)}}{( 1- e^{-\beta E_0} )^2}   \, .
    \label{eq:Q}
\end{eqnarray}
We test this approximation in figure~\ref{fig:transition-entropy} for a sample network with $N=105$ nodes and find a very good agreement with the numerically exact values.

For large networks and low effective temperatures, we can further simplify expression \eqref{eq:Q} by noting that $e^{-\beta E_0 (M+1)} / e^{-\beta E_0} \rightarrow 0$ for $\beta \rightarrow \infty$ or in large networks $M \rightarrow \infty$. 
We then obtain
\begin{equation}
    Q_{\rm st}  \sim
    \frac{1}{(1- e^{-\beta E_0} )^{2}}   \, .
    \label{eq:Q-high-beta}
\end{equation}
In this limit, the entropy can be computed from equation~\eqref{eq:H-from-Q} as
\begin{equation}
    H_{\rm st} \sim -2 \ln (1-e^{-\beta E_0})   
    + \frac{2\beta E_0 e^{-\beta E_0}}{
    1-e^{-\beta E_0}} \, .
    \label{eq:H-high-beta}
\end{equation}
We find that the entropy vanishes for low effective temperatures, 
\begin{equation*}
    H_{\rm st} \rightarrow 0
    \quad \mbox{for} \quad
    \beta \rightarrow \infty,
\end{equation*}
indicating that all nodes $i$ choose a single supplier. 
In fact, the production is local in this limit. 

The entropy differs substantially from zero when $\beta E_0$ is of the order of unity leading to the estimate $\beta_{\rm crit} \approx E_0^{-1}$ for the critical value. 
A slightly more accurate approximation can be obtained by computing the value for which the entropy $H$ equals half of its maximum value which yields the implicit condition
\begin{equation}
    H_{\rm st}(\beta_{\rm crit}) = \frac{1}{2} \ln(N). 
    \label{eq:Hst-half-max}
\end{equation}
For higher values of the effective temperature (lower values of $\beta$), the approximation \eqref{eq:H-high-beta} is no longer valid as indicated in figure~\ref{fig:transition-entropy}

\subsection{From centralized to diverse production}

We finally consider the transition between the diverse (III) and the centralized production regime (II). This transition is driven by the competition of the two terms in the agents' utility function \eqref{eq:utility-agent} -- the individual differences and the universal prices. A diverse production is found if the preferences are sufficiently different, i.e.~if the weight parameter (the effective temperature) $\TT$ exceeds a critical value $\TT_{\rm crit}$, cf.~figure~\ref{fig:phase_diagrams}. 

\begin{figure}[tb]
    \centering
    \includegraphics[trim = 0cm 8cm 0cm 0cm, width=.75\columnwidth]{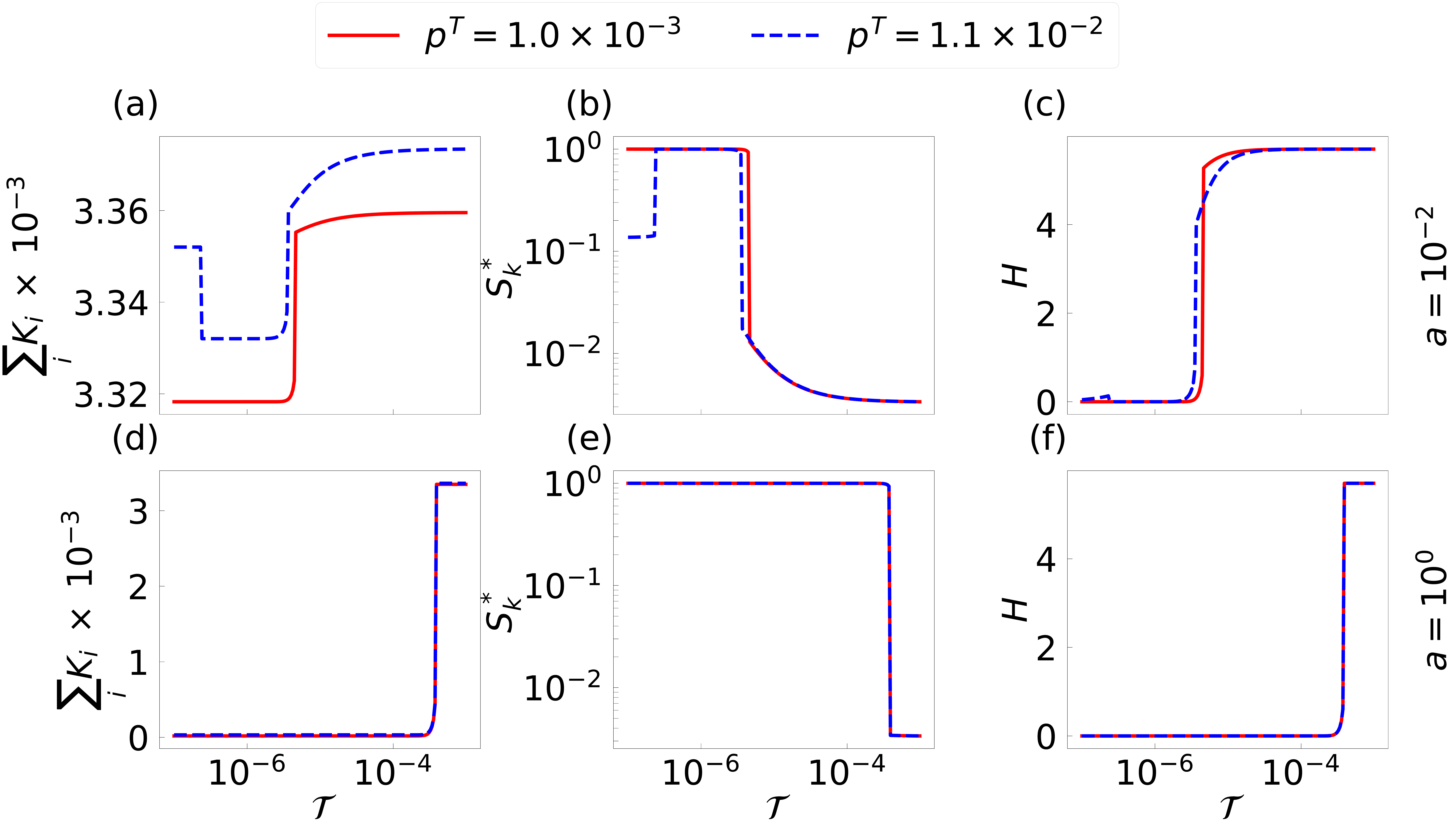}
    \caption{
    Transition from diverse to centralized production. 
    Results are shown for a single instance of the transportation network, comparing the case of (a-c) weak scale effects and (d-f) strong scale effects as well as different values of the specific transportation costs $p^T$ (red solid vs.~blue dashed line). 
    We show (a,d) the total costs $\sum_i K_i$, (b,e) the production of the busiest node $S^*_k = \max_k S_k$ and (c,f) the entropy $H_i$ averaged over all nodes.
    The transition is very sharp and occurs at a critical value of the effective temperature given by equation~\eqref{eq:23-Tcrit}. A non-monotonous behavior of $S_k^*$ is found for  $a=10^{0}$ and $p^T=1.1\times 10^{-2}$, which is discussed in the text.
    }
    \label{fig:transition-II-III-single}
\end{figure}

This critical value $\TT_{\rm crit}$ can be estimated in terms of the aggregated interpretation introduced in section \ref{sec:aggregated} by comparing the different contributions to the free energy \eqref{eq:utility} in the two regimes. 
If production is fully diversified, then the supply matrix is given by $S_{ji} = D/N$ (cf.~equation.~\eqref{eq:Smat-diverse}) and the free energy  \eqref{eq:utility} is thus given by
\begin{eqnarray*}
    \KK_{i, \rm div} &\approx (\bar b - a D) D + p^T \bar E_i D\\
    \Rightarrow  \FF_{i, \rm div} &\approx \TT \ln(N) - \KK_{i, \rm div},
\end{eqnarray*}
where $\bar E_i$ is the average distance from node $i$ to all other nodes, $\bar E_i = N^{-1} \sum_{j} E_{ij}$ and $\bar b = N^{-1} \sum_{j} b_{j}$.
If production is fully centralized at a node $n$, then the supply matrix is given by $S_{ji} = D \delta_{jn}$ using the Kronecker delta symbol (cf.~equation~\eqref{eq:Smat-central}). 
In this case, the free energy  \eqref{eq:utility} reads
\begin{eqnarray*}
    \KK_{i, \rm cen} &\approx (b_n - a N D) D + p^T E_{in} D \\
    \Rightarrow \FF_{i, \rm cen} &\approx - K_{i, \rm cen} \, .
\end{eqnarray*}
We expect the production to be purely centralized when $\FF_{i, \rm cen}  > \FF_{i, \rm div}$ for all nodes $i$ and to be diverse if $\FF_{i, \rm div} > \FF_{i, \rm cen}$ for all $i$. 
If scale effects are sufficiently strong, then the differences between the nodes $i$ are negligible and the transition is abrupt. 
We then expect the transition to take place at a critical temperature $\TT_{\rm crit}$ given by 
\begin{eqnarray*}
    & \FF_{i, central}(\TT_{\rm crit}) \approx \FF_{i, diverse}(\TT_{\rm crit}), \\
    \Rightarrow & \TT_{\rm crit} \approx \frac{\KK_{i, \rm div} - \KK_{i, \rm cen}}{\ln(N)}.
\end{eqnarray*}
Neglecting the inhomogeneities in the $b_i$ and the transportation distances, we finally obtain  
\begin{equation}
    \TT_{\rm crit} \approx \frac{a (N-1) D^2}{\ln(N)}. 
    \label{eq:23-Tcrit}
\end{equation}
We conclude that the transition between diverse and centralized production is driven by the competition of scale effects scaling linearly in $a$ and diversity in preferences scaling linearly in $\TT$, while transportation effects play a negligible role.  Hence, the critical effective temperature $\TT_{\rm crit}$ scales linearly in $a$, while it is largely independent of the specific transportation costs $p^T$. However, this reasoning is only valid as long as local production is not competitive, i.e.~as long as $p^T$ is small enough.

A surprising behavior is found for weak scale effects and intermediate values of the specific transportation costs $p^T$ (figure~\ref{fig:transition-II-III-single}, upper row, dashed line). 
We find that the maximum production $S^*$ jumps to one when the effective temperature $\TT$ increases above approximately $10^{-6}$. 
That is, an increase in the diversity of preferences leads to a decrease in product diversification. This counter-intuitive behavior is a consequence of the multistability of the economic system. 
Two Nash equilibria exist for $\TT \ll 10^{-6}$, one with an incomplete and one with a complete centralization of production. 
Due to the design of our numerical experiments the initial state features an incomplete centralization. 
An increase in $\TT$ fosters trade, strengthening one producer at the expense of another one. 
Eventually, the Nash equilibrium with incomplete centralization becomes unstable, and the system relaxes to the centralized equilibrium, offering lower total costs due to scale effects. 
Notably, a further increase of the effective temperature finally leads to a fully diversified production as expected.

\subsection{Comparison to numeric results}
\label{sec:compare_theo_numerics}

\begin{figure}[tb]
    \centering
    \includegraphics[trim = 0cm 0cm 0cm 0cm, width=.75\columnwidth]{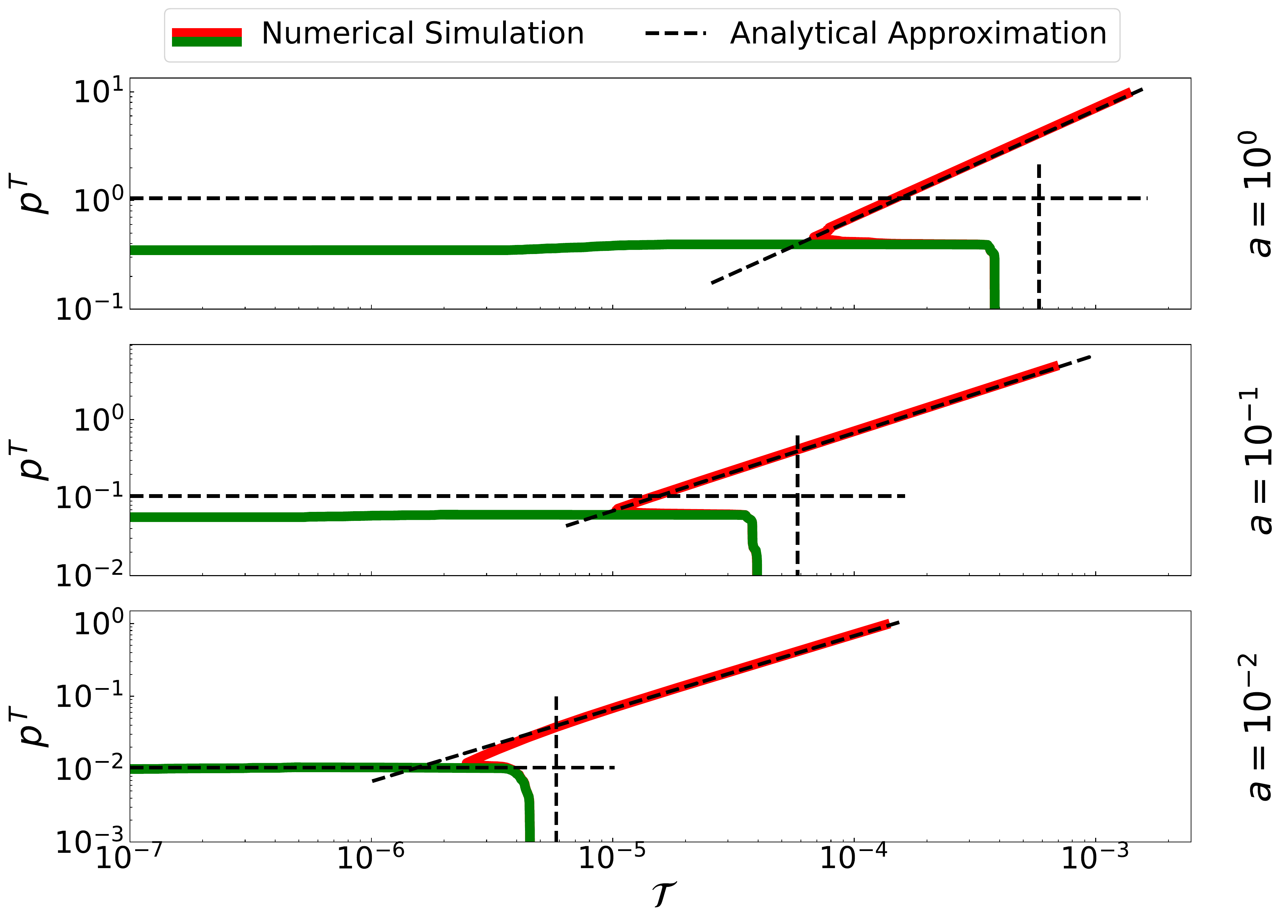}
    \caption{
    Comparison of numerical results and analytic estimates for the location of phase transitions in terms of the parameters $p^T$ and $\TT$. Green and red lines show the numerical results as established in figure~\ref{fig:phase_diagrams}. 
    Black dashed lines show the corresponding analytical estimates according to equations~\eqref{eq:ptcrit-I-II-a-large}, \eqref{eq:Hst-half-max}, and \eqref{eq:23-Tcrit}. 
    The strength of the scale effects is varied as $a=10^0$ (top panel), $a=10^{-1}$ (middle panel) and $a=10^{-2}$ (bottom panel).
    }
    \label{fig:analytical_comp}
\end{figure}

As a final step of our analysis, we compare the analytical estimates for the location of the phase boundaries to the numerical results in figure~\ref{fig:analytical_comp}. 
We find that the estimate \eqref{eq:Hst-half-max} for the phase boundary between the localized and diversified phase accurately matches the numerical results with no visible differences. 
The estimate \eqref{eq:23-Tcrit} slightly overestimates the critical value $\TT_{\rm crit}$ for the transition from centralized to diversified production. 
However, the analytical estimate faithfully reproduces the order of magnitude of $\TT_{\rm crit}$ as well as the scaling with the parameter $a$.
Similarly, the estimate \eqref{eq:ptcrit-I-II-a-large} provides a fair overall estimate for the critical value $p^T_{\rm crit}$ for the transition from localized to centralized production. 
The scaling with $a$ is overestimated such that analytical estimates are larger than the numerically exact values for large $a$. 
This is not surprising as we expect a smooth crossover between the two different scaling regimes. 
For medium to large values of $a$, we expect a proportional scaling of $p^T_{\rm crit}$ with $a$ as described by \eqref{eq:ptcrit-I-II-a-large}. 
In contrast, $p^T_{\rm crit}$ should tend to a constant for small values of $a$ as there is still a (continuous) centralization at a finite $p^T_{\rm crit}$ due to the small inhomogeneity in the $b_i$.

\section{Conclusion and outlook}

In this article, we have established a model for the formation of trade networks based on the decisions of economic agents. 
The model combines two driving factors for the emergence of trade: On the supply side, trade is established if regional differences in production costs, including economic scale effects, exceed the transportation costs.
On the demand side, the diversity of the agents' preferences fosters trading even if this increases costs. 

The model was derived from the decisions of individual agents in terms of discrete choice theory. Every agent decides for a supplier based on both the price and an individual factor modeled as a random variable. On an aggregated scale, a set of agents thus purchases goods from different suppliers at different locations. On this aggregated level, the model shows strong connections to ensembles in statistical thermodynamics as the equilibrium on the agent level coincides with an equilibrium of the effective free energy \eqref{eq:utility}.
However, all decisions are coupled through scale effects: The decision of any agent changes the prices for all other agents and may thus trigger further decisions. 
Hence, the common equilibrium of statistical physics must be generalized to a Nash equilibrium. 
Nevertheless, statistical thermodynamics provides essential concepts to compute the equilibrium states and to understand the emergence of a trade network.

We have shown that the model bears three different regimes of trade. 
If transportation costs are high and the diversity of preferences is weak, then all goods are produced locally. 
A trade network can emerge in two different ways rooted in either the supply or demand side. 
Decreasing transportation costs makes it cheaper to import goods thus fostering the emergence of trade. This process is essentially driven by economies of scale, as every increase of production leads to lower prices fostering further increase in exports.
Eventually, this process leads to a complete centralization of production and a directed star-like trade network. 
An increase in the diversity of preferences drives the emergence of bilateral trade and eventually leads to an all-to-all coupled trade network. We have developed a comprehensive analytical theory of the transitions between these regimes and derived analytical estimates for critical parameter values. 
Remarkably, the transition to a phase with centralized production can be continuous as well as discontinuous if economic scale effects are strong enough.

The model describes some essential mechanisms in the formation of trade networks -- but of course it cannot capture all facets of this complex process. For instance, our model captures the importance of economies of scale, diverse preferences and emergent hysteresis effects \cite{krugman1981trade,schroder2018hysteretic}.
A limitation of the model is found in the emerging production patterns: When transportation costs continue to decrease, production will either be centralized completely or not at all depending on the parameter $\TT{}$.
In reality, one often observes multi-center structures, for example in urban systems \cite{fujita1982multiple,louf2013modeling} as a result of spatial constraints or congestion at high levels of centralization. 
In the proposed modeling framework, this would be possible if scale effects saturated or if transportation capacities were limited. 

From a more fundamental viewpoint, the model starts from discrete choice theory and then aggregates over many agents to obtain the total purchases between the nodes of a network. Other economic models, such as the celebrated Dixit-Stiglitz model maximize utility for a given budget \cite{dixit1977monopolistic}.
From a statistical viewpoint, a thorough treatment of the thermodynamic limit remains challenging due to the different scaling behaviors in the utility function. 
This analysis is well beyond the scope of the present analysis, so we have focused on finite systems.

\appendix

\section{Proof of discontinuous transition}
\label{sec:proof-disc}

In this appendix we proof the existence of a discontinuous transition in a single step from Sec.~\ref{sec:transition-I-III}. We have $\TT{} = 0$ such that all agents at a node $i$ purchase from same node $i^*$. We can thus treat a node as a single entity in the following analysis. 

We first consider the purchases of node $i=2$, that may choose to purchase from node $j=1$ or produce locally. By an explicit computation, we can show that local production ($S_{22} = D$, $S_{12} = 0$) is cheaper for $p^T > p^T_{\rm crit}$. Importing goods ($S_{22} = 0$, $S_{12} = D$) is cheaper for $p^T < p^T_{\rm crit}$. 

Due to the ordering of the nodes we can conclude two further statements: (i) Node $i=2$ will change its purchases to buy at node $j=1$, while purchasing at another node $\ell \ge 3$ provides no benefits. (ii) Node $i=2$ is the first one to change its purchases. That is, production is fully locally ($S_{\ell,\ell} = D$ for all nodes $\ell=1,\ldots,N$) for $p^T > p^T_{\rm crit}$.

Now consider the consequences of node $i=2$ changing its purchases at  $p^T = p^T_{\rm crit}$. Node $i'=3$ can either buy locally at a price
\begin{equation}
    p_{33} = b_3 - a D 
\end{equation}
or at node $j=1$ at a price
\begin{equation}
    p_{13} = b_1 - 3 a D + p^T_{\rm crit} \, E_{13} \, .  
\end{equation}
By our assumption \eqref{eq:centralization-discontinuous-a}, we find that $p_{13} < p_{33} $ such that node $i'=3$ changes its purchases from local to import simultaneously with node $i=2$. The same argument now applies to all nodes $i''=4\ldots,N$. Hence, we conclude that at $p^T = p^T_{\rm crit}$ all nodes simultaneously change their purchases such that we find
\begin{eqnarray}
    \mbox{for} \, p^T > p^T_{\rm crit}: & \quad S_{ii} = D, \nonumber \\
    \mbox{for} \, p^T < p^T_{\rm crit}: & \quad S_{1i} = D,
\end{eqnarray}
for all nodes $i=1\ldots,N$.

\section{Entropy and partition function}
\label{sec:app-entropy}

In this appendix we briefly recall the relation of the entropy and the partition function~\eqref{eq:partition}. In particular, we show that the expression \eqref{eq:H-from-Q} for the entropy $H_i$ is equivalent to the definition \eqref{eq:entropypurchase}. Differentiating the partition function $Q_i$ with respect to $\beta$ yields
\begin{eqnarray}
    -\frac{\partial}{\partial \beta} \ln(Q_i) 
    &= -\frac{1}{\sum_{l =1}^N e^{-\beta E_{li}} } \,\frac{\partial}{\partial \beta} \sum_{j =1}^N e^{-\beta E_{ji}}, \nonumber \\
    &= \sum_{j =1}^N \frac{e^{-\beta E_{ji}}}{\sum_{l =1}^N e^{-\beta E_{li}}}\,E_{ji}, \nonumber \\
    &= \sum_{j =1}^N \frac{S_{ji}}{D} E_{ji}.
\end{eqnarray}
Similarly, we can show that
\begin{eqnarray}
    -\frac{\partial}{\partial \beta^{-1}}(-\beta^{-1}\ln{Q_i}) 
    &= \ln{Q_i} - \beta \frac{\partial}{\partial \beta} \ln{Q_i} \nonumber \\
    &= \ln{Q_i} + \beta \sum_{j =1}^N \frac{S_{ji}}{D} E_{ji} .
    \label{eq:app-entropy-Qdiff}
\end{eqnarray}
Next, we start from the definition of entropy,
\begin{eqnarray}
    H_i &= -\sum_{j=1}^N \frac{S_{ji}}{D}\, \ln{\frac{S_{ji}}{D}} \nonumber \\
        &= -\sum_{j=1}^N \frac{e^{-\beta E_{ji}}}{\sum_{l =1}^N e^{-\beta E_{li}}}\, 
        \ln{\frac{e^{-\beta E_{ji}}}{\sum_{l =1}^N e^{-\beta E_{li}}}} \nonumber \\
        &= -\sum_{j=1}^N \frac{e^{-\beta E_{ji}}}{\sum_{l =1}^N e^{-\beta E_{li}}}
        \left[ \ln{e^{-\beta E_{ji}}} - \ln{\sum_{l =1}^N e^{-\beta E_{li}}} \right] \nonumber \\
        &= \sum_{j=1}^N \frac{e^{-\beta E_{ji}}}{\sum_{l =1}^N e^{-\beta E_{li}}}\, \beta \, E_{ji}
        + \ln{ \left( \sum_{l =1}^N e^{-\beta E_{li}} \right) }\, \sum_{j=1}^N \frac{e^{-\beta E_{ji}}}{\sum_{l =1}^N e^{-\beta E_{li}}} \nonumber \\
        &= \beta \, \sum_{j=1}^N \frac{S_{ji}}{D}\,  E_{ji} + \ln{(Q_i)},
\end{eqnarray}
which coincides with the expression \eqref{eq:app-entropy-Qdiff}.

\ack

We thank Joahnnes Többen for stimulating discussions. We gratefully acknowledge support from the German Federal Ministry of Education and Research (BMBF) via the grant \textit{CoNDyNet2} with grant no. 03EK3055B, the Helmholtz Association via the grant \textit{Uncertainty Quantification -- From Data to Reliable Knowledge (UQ)} with grant no.~ZT-I-0029, and the Deutsche Forschungsgemeinschaft (DFG, German Research Foundation) with grant No. 491111487.

\section*{Data availability statement}
The data that support the findings of this study are openly available~\cite{philipp_c_bottcher_2023_7628524}.

\section*{References}

\providecommand{\newblock}{}

\end{document}